\documentclass[useAMS,usenatbib]{mn2e}
\usepackage{aas_macros,graphicx,times}
\usepackage{multirow}
\title[X-ray observations of \src]
  {Unveiling the nature of \src\ with \xmm\thanks{Based on observations obtained with XMM-Newton, an ESA science mission with instruments and contributions directly funded by ESA Member States and NASA.}}
\author[P. Esposito et al.]
  {P.~Esposito,$^{1,2}$\thanks{E-mail: paoloesp@iasf-milano.inaf.it}
A.~De~Luca,$^{2,3}$ A.~Tiengo,$^2$ A.~Paizis,$^2$ S.~Mereghetti,$^2$ and P.~A.~Caraveo$^2$
\smallskip\\ 
 $^1$Universit\`a degli Studi di Pavia, Dipartimento di Fisica Nucleare e Teorica and INFN-Pavia, Pavia, Italy\\
  $^2$INAF - Istituto di Astrofisica Spaziale e Fisica Cosmica, Sezione di Milano, Milano, Italy\\
$^3$IUSS - Istituto Universitario di Studi Superiori, Pavia, Italy
}
\date{Accepted 2007 November 05. Received 2007 November 05; in original form 2007 October 29}

\pagerange{\pageref{firstpage}--\pageref{lastpage}} \pubyear{2007}

\def\LaTeX{L\kern-.36em\raise.3ex\hbox{a}\kern-.15em
    T\kern-.1667em\lower.7ex\hbox{E}\kern-.125emX}

\def\src {RX\,J0002+6246}
\def\tmsrc {2MASS\,00025569+6246175}
\def\xmm {\emph{XMM-Newton}}

\def\int {\emph{INTEGRAL}}

\def\rst {\emph{ROSAT}}

\def\flux {\mbox{erg cm$^{-2}$ s$^{-1}$}}
\def\lum {\mbox{erg s$^{-1}$}}
\def\nh {$N_{\rm H}$}

\begin{document}

\label{firstpage}

\maketitle

\begin{abstract}
The X-ray source \src\ was discovered close to the supernova remnant CTB\,1 in a \rst\ observation performed in 1992. The source phenomenology (soft spectrum, apparent lack of counterparts, possible pulsations at 242 ms, hints for surrounding diffuse emission) led to interpret it as an isolated neutron star in a new supernova remnant. We have analysed an archival \xmm\ observation performed in 2001. The source coordinates, as computed on the \xmm\ images, coincide with those of a bright source listed in optical and infrared catalogues. The X-ray spectrum is well described by an optically thin plasma model. No fast pulsations are seen, nor clear evidence of a supernova remnant associated to the source. Thus, we conclude that \src\ is not an isolated neutron star, but the X-ray counterpart of the bright optical/infrared source, most likely a F7 spectral class star located at about $0.2$ kpc.

\end{abstract}


\section{Introduction}
Most of the observed isolated neutron stars are identified as pulsars, whose emission derives either from rotational energy loss, or as magnetars, powered by magnetic field decay \citep[e.g.,][]{manchester04}. The central compact objects \citep[CCOs; see][]{pavlov02} remain perhaps the least understood members of the isolated neutron stars family. These X-ray sources are located within supernova remnants (SNRs) and, despite intensive campaigns, have not been detected as radio or optical sources so far. CCOs are seemingly young (\mbox{$\la$10$^4$ years}) isolated neutron stars, with steady X-ray fluxes (with the notably exception of 1E\,161348$-$5055 at the centre of the SNR RCW\,103; \citealt{deluca06}), soft thermal spectra, and lack of surrounding pulsar wind nebulae. Two CCOs out of seven are pulsating sources: 1E\,1207.4$-$5209 in G\,296.5+10.0, with period \mbox{$P=424$ ms} \citep{zavlin00}, and PSR\,J1852+0040 in Kes\,79, with \mbox{$P=105$ ms} \citep{ghs05}. Both sources have small spin-down rates with period derivatives $\dot{P}<2\times10^{-16}$ s s$^{-1}$ \citep{gotthelf07,halpern07}.\\
\indent The current sample of CCOs includes seven ``confirmed'' sources and four ``candidates''. The confirmed CCOs are the central sources in RCW~103 \citep{tuohy80}, G\,296.5+10.0 \citep{helfand84}, Pup~A \citep{petre96}, Vela~Jr. \citep{aschenbach98}, G\,347.3$-$0.5 \citep{slane99}, Cas~A \citep{tananbaum99}, and Kes~79 \citep{seward03}. The candidates are those in G\,349.7+0.2 \citep{lazendic05}, G\,15.9+0.2 \citep{reynolds06cco}, G\,330.2+1.0 \citep{park06}, and \src\ in G\,117.9+0.6 \citep{hailey95}, which is the object of our research.\\
\indent The X-ray point source \src\ was discovered with the PSPC instrument on board \rst\ near the supernova remnant CTB\,1, during a \mbox{$\sim$9 ks} long observation carried out on 1992 August 16--17.
\citet{hailey95} reported the position \mbox{$\rm{R.A.}=00^{\rm{h}}02^{\rm{m}}54\fs1$}, $\rm{Decl.}=62^{\circ}46'23''$ (epoch J2000).
The observation showed a hint of a faint shell of soft X-ray emission (G\,117.7+0.6), proposed as a SNR associated with \src\ \citep{hailey95,craig97}. 
The spectrum of the point source was fitted using a blackbody attenuated by interstellar absorption, with \mbox{$k_BT\simeq0.15$ keV}. Assuming a distance of 3 kpc \citep{hailey95}, this corresponds to a 0.5--2 keV luminosity of \mbox{$\sim$$2\times10^{32}d_3^2$ \lum} (where we indicate with $d_{\rm{N}}$ the distance in units of N kpc). \citet{hailey95} also found some evidence for a possible periodicity in the X-ray emission of \src\ with period \mbox{$P=242$ ms}. Based on these results,  \citet{hailey95} proposed that \src\ is an isolated neutron star in a SNR. Furthermore, the absence of counterparts at other wavelengths \citep[][]{hailey95,brazier99} suggested that \src\ could be a CCO.\\
\indent In 2001 \src\ has been observed for 33 ks with \xmm. \citet{pavlov04} reported results of the analysis of those data. The spectrum was fitted by a two-component model: a soft blackbody with temperature \mbox{$k_BT\simeq0.1$ keV} and a hard component, either a second blackbody with $k_BT\simeq0.5$ keV or a power-law with photon index $\Gamma\simeq2.6$. Any periodicity was excluded, as well as the presence of a SNR around the source. They concluded that \src\ is most likely a middle-aged pulsar rather than a CCO.\\
\indent Here we report on a re-analysis of the \xmm\ observation of \src. This paper presents evidence that, contrary to previous claims, the X-ray source \src\ is neither a CCO nor a pulsar, but rather a non-degenerate star. We have also identified its likely stellar counterpart using near-infrared data.
\section{Observation and analysis}
The \xmm\ X-ray observatory observed the field of \src\ for 33 ks in 2001, from August 22 17:16 UT to August 23 02:30 UT (observation ID: 0016140101). The data were collected with the EPIC instrument, which consists of two MOS \citep{turner01short} and one pn \citep{struder01short} cameras sensitive to photons with energies between 0.1 and 15 keV. 
The EPIC pn was operated in Small Window mode (time resolution 6 ms) while the EPIC MOS had the MOS\,1 and MOS\,2 units in Full Frame mode (time resolution \mbox{2.6 s}). Both the pn and MOS mounted the medium thickness filter.\\
\indent  All the data reduction was performed using the Science Analysis Software (\textsc{SAS}) software package\footnote{See \mbox{http://xmm.vilspa.esa.es/}.}  version 7.1. The raw observation data files were processed using standard pipeline tasks (\textsc{epproc} for pn, \textsc{emproc} for MOS data).\\
\indent We selected events with pattern 0--4 and pattern 0--12 for the pn and the MOS, respectively. To obtain the results presented in this work we filtered the data to reject intervals with soft-proton flares, reducing the net exposure time to 10.4 ks for the pn detector,  \mbox{18.3 ks} for the MOS\,1, and 18.5 ks for the MOS\,2.

\subsection{Spatial analysis}\label{imaging}
For the imaging analysis we used the EPIC MOS data, since the pn camera in Small Window mode covers only a $4'\times4'$ sky region, while MOS cameras were exposed in Full Frame mode, providing a 30$\arcmin$ diameter field of view. 
\subsubsection{Absolute astrometry}
\begin{figure}
\centering
\resizebox{\hsize}{!}{\includegraphics[angle=0]{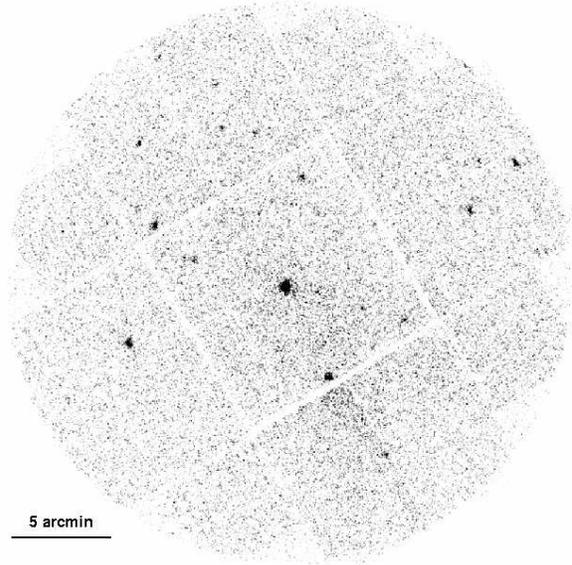}}
\caption{\xmm\ EPIC MOS image of the field of \src\ in the 0.3--2 keV energy range. North is to the top, east to the left. The imagine has been smoothed with a Gaussian function with kernel radius of three. The image shows the hint of faint diffuse emission (to the south-west) discussed in Sect. \ref{diffuse}.}
\label{xmm-fov}
\end{figure}
The brightest point source in the 0.3--2 keV image (Fig.~\ref{xmm-fov}) is detected near the centre of the field of view. The \mbox{\textsc{emldetect}} routine reports a best-fit position of $\rm{R.A.}=00^{\rm{h}}02^{\rm{m}}55\fs8$, \mbox{$\rm{Decl.}=62^{\circ}46'17\farcs9$} (epoch J2000), with an uncertainty of $0\farcs2$. This (1\,$\sigma$) uncertainty is statistical and does not include the systematic uncertainty in \xmm\ pointing. Given the brightness of the source, the statistical error is smaller than the absolute astrometric accuracy of \xmm\  \citep[$1\farcs5$ root mean square;][]{kirsch04}.\\
\indent We have detected about 40 point sources within the total field of view, most of them without an obvious counterpart at other wavelengths. We measured the position of the bright star TYC\,4018--2777--1, visible in X-rays, to be $1\farcs4$ from its USNO-B1.0 catalog\footnote{See http://www.nofs.navy.mil/data/fchpix/.} \citep[][]{monet03short} position, entirely consistent with the expected EPIC astrometric accuracy (the systematic uncertainty in connecting the USNO astrometry to the International Celestial Reference System is $0\farcs2$ in each coordinate). The lack of other X-ray sources with clear optical/infrared identification does not allow us to unambiguously register the X-ray image on the optical plates.\\
\indent With respect to the nominal \rst/PSPC position of \src\ reported by the WGACAT\footnote{See http://wgacat.gsfc.nasa.gov/wgacat/wgacat.html\,.} Rev.~1 \citep{white94} and \citet{hailey95}, the positional offset of the \xmm\ source is 12.6$\arcsec$. The astrometric accuracy of the WGACAT catalog is roughly 13$''$ (1$\sigma$ error). The source positions are then well consistent within the uncertainties. Since no other X-ray source is consistent with the \rst\ position of \src, here and in the subsequent discussion we assume that the source detected in the EPIC cameras and \src\ are the same X-ray source.
\subsubsection{Identification of infrared counterparts}\label{ircounterpart}
We searched for optical or infrared counterparts of \src\ around our best-fit position in various catalogs, including the Two Micron All Sky Survey\footnote{See  http://www.ipac.caltech.edu/2mass/.}  \citep[2MASS;][]{skurtsie06short}. The 2MASS database covers the entire sky and its Point Source Catalog gives the positions and J (\mbox{1.25 $\mu$m}), H (1.65 $\mu$m), and K$_{\rm{s}}$ (\mbox{2.17 $\mu$m}) magnitudes of its sources. The astrometric accuracy of this catalog is better than $0\farcs1$.\\
\indent The only object from the 2MASS catalog with a position inside the \xmm\ error circle is \tmsrc. This source lies at \mbox{$\rm{R.A.} = 00^{\rm{h}}02^{\rm{m}}55\fs70$}, \mbox{$\rm{Decl.} = 62^{\circ}46'17\farcs6$} (epoch J2000), only $0\farcs6$ from the centroid of the X-ray source. Its magnitudes are $10.32\pm0.02$, $9.94\pm0.03$, and $9.81\pm0.03$ in the J, H, and  K$_{\rm{s}}$ bands, respectively. The random chance probability of finding an object as bright in the near-infrared as \tmsrc\ (or brighter) inside the \xmm\ error circle (at a 99\% confidence level) is smaller than $2\times10^{-5}$, making the association with \src\ very likely. The second closest infrared source to \src\ lies at more than $10\arcsec$ from its X-ray position.
\subsubsection{Diffuse X-ray emission}\label{diffuse}
The EPIC images hint the existence of a faint structure of diffuse emission located to the South-West of \src. Its surface brightness is of $(7\pm2)\times10^{-4}$ counts s$^{-1}$ arcmin$^{-2}$ in the \mbox{0.3--2 keV} energy range. A detailed spectral analysis of such a faint feature is hampered by the low signal-to-noise ratio.\\
\indent This diffuse structure, also detected in \rst\ images, as well as in radio maps \citep[][and references therein]{craig97}, corresponds to an apparent North-East extension of the nearby SNR CTB1. Although the nature of such diffuse emission remains unclear (it could be related to CTB1, or have a different origin -- the region is complex and permeated by several diffuse features), it is most likely unrelated to \src.

\subsection{Timing analysis}\label{timing}
We searched for pulsed X-ray emission from \src\ using the high time resolution pn data (6 ms time resolution). Source photons were selected in the \mbox{0.3--2 keV} energy range from a circular region centred on \src\ with radius of $30''$. Photon arrival times were converted to the solar system barycentre using the SAS task \textsc{barycen}. For the barycentric correction, we used the position inferred from the MOS image fitting (see Sect.~\ref{imaging}).\\
\indent We searched the data for pulsations using the $Z^2_n$ test \citep{buccheri83}, with the number of harmonics $n$ being varied from 1 to 4. We searched for a pulsed signal over a wide period range centred on the value suggested by \citet{hailey95} ($0.24181\pm0.00001$ s). No statistically significant signal was detected. We found a 99\% confidence upper limit on the pulsed fraction of 15 percent (assuming a sinusoidal modulation). Indeed, also the detection of the modulation reported by \citet{hailey95} was marginal. We then searched the data for pulsations to a minimum period of 12 ms, but we again did not detect any significant signal.

\subsection{Spectral analysis}
\begin{figure}
\centering
\resizebox{\hsize}{!}{\includegraphics[angle=-90]{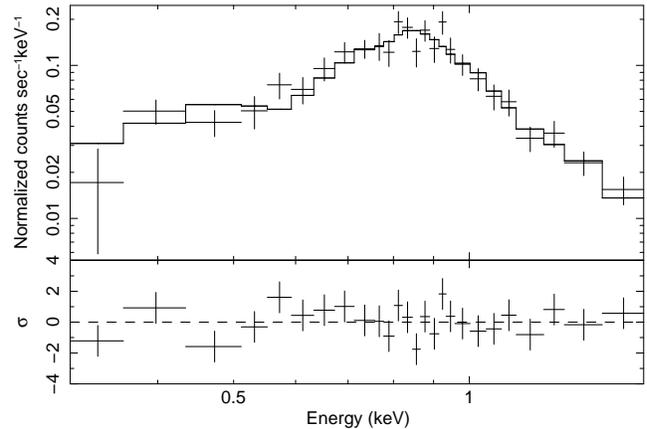}}
\caption{EPIC pn spectrum of \src. Upper panel: data and best-fit \textsc{MEKAL} model for the parameters given in Table~\ref{fits}. Lower panel: residuals in units of sigma.}
\label{spec}
\end{figure}

The source spectra were accumulated from circular regions ($30''$ radius) centred on \src. The background spectra were extracted from source-free regions of the same chip as the source: annular regions with radii of 80$\arcsec$ and 125$\arcsec$ for the MOSs, and a rectangular region with area of \mbox{$\sim$$2.8\times10^3$ arcsec$^2$} located on the side of the source for the pn. We carefully checked that the choice of different background extraction regions did not affect the spectral results. During the observation, between 0.3 and 2 keV a total of $905\pm32$ counts above the background were collected from \src\ by the pn detector, $400\pm21$ by the MOS\,1 detector, and $422\pm22$ by the MOS\,2 detector.\\
\indent Spectral redistribution matrices and ancillary response files were generated using the \textsc{SAS} scripts \textsc{rmfgen} and \textsc{arfgen}, and spectra grouped  with a minimum of 30 counts per energy bin were fed into the spectral fitting package \textsc{XSPEC}\footnote{See \mbox{http://heasarc.gsfc.nasa.gov/docs/xanadu/xspec/}.} version 12.3. Spectral channels having energies below 0.3 keV and above 2.0 keV were ignored, owing to the very low counts from \src.\\
\indent We jointly fit the spectra by MOS\,1, MOS\,2, and pn to a number of different models including a blackbody, power-law, blackbody plus power-law, two blackbodies, bremsstrahlung, Raymond-Smith plasma \citep{raymond77},  \textsc{MEKAL} \citep{mewe85,mewe86,liedahl95}, and \textsc{APEC} \citep{smith01}, all corrected for interstellar absorption. The abundances used are those of \citet{anders89} and photoelectric absorption cross-sections from \citet{balucinska92}. The data are well described by the Raymond-Smith, \textsc{MEKAL}, and \textsc{APEC} models, with plasma temperatures of $\sim$0.7 keV (see Table~\ref{fits} for the best-fit model parameters), whereas all the other models yield statistically unacceptable fits (with $\chi_r^2>1.5$). In Fig.~\ref{spec} the spectrum of \src\ fitted with the \textsc{MEKAL} model is shown.
\begin{table*}
\begin{minipage}{15cm}
\centering
\caption{Spectral results in the 0.3--2 keV energy range. Errors are quoted at the 90\% confidence level for a single interesting parameter.}
\label{fits}
\begin{tabular}{@{}ccccccc}
\hline
Model$^{\mathrm{a}}$ & \nh & $k_BT$ & Metal abundances$^{\mathrm{b}}$ & Absorbed flux$^{\mathrm{c}}$ & Unabsorbed flux$^{\mathrm{c}}$ & $\chi_r^2$ (d.o.f.)\\ 
 &  ($10^{21}$ $\rm cm^{-2}$) & (keV) & & (\flux) & (\flux) & \\
\hline
\textsc{RS} & $1.1\pm0.3$ &  $0.75\pm0.05$ &  $0.09^{+0.05}_{-0.03}$ &  $1.3\times10^{-13}$ & $2.4\times10^{-13}$ & 1.18 (49)\\ 
\textsc{APEC} & $1.4^{+0.3}_{-0.4}$ & $0.66^{+0.07}_{-0.04}$ & $0.10^{+0.03}_{-0.02}$ &  $1.3\times10^{-13}$ & $2.6\times10^{-13}$ & 1.00 (49)\\
\textsc{MEKAL} & $1.5\pm0.3$ & $0.64\pm0.04$ & $0.09\pm0.02$ & $1.3\times10^{-13}$ & $2.8\times10^{-13}$ & 0.93 (49)\\
\hline
\end{tabular}
\medskip
\begin{list}{}{}
\item[$^{\mathrm{a}}$] Models applied in \textsc{XSPEC} notation:  \mbox{\textsc{RS = phabs*raymond}}, \textsc{APEC = phabs*apec}, and \mbox{\textsc{MEKAL = phabs*mekal}}.
\item[$^{\mathrm{b}}$] With the abundance ratios of \citet{anders89}.
\item[$^{\mathrm{c}}$] Flux in the 0.3--2 keV energy range.
\end{list}
\end{minipage}
\end{table*}

\section{Discussion and Conclusions}
The X-ray source \src\ is clearly detected in the \xmm\ images and its position is consistent with a rather bright star (\tmsrc).  
\citet{hailey95} ruled out this star as a possible counterpart of \src\ mainly for the angular separation of $12\arcsec$ from their X-ray position. However, they relied upon a positional uncertainty of $10\arcsec$, a value that in subsequent releases of the WGACAT was conservatively increased to $13\arcsec$ (1\,$\sigma$). Moreover, the source coordinates in the WGACAT are affected by a systematic error\footnote{See Haberl F., Pietsch W., and Voges W., ``Differences in the two ROSAT catalogs of pointed PSPC observations'' (1994), and comments by White N. E., Angelini L., and Giommi P.. The document is available at ftp://ftp.xray.mpe.mpg.de/rosat/catalogues/sourcecat/wga\_rosatsrc.html\,.}. The recent (2001) Second ROSAT Source Catalog of Pointed Observations with the Position Sensitive Proportional Counter\footnote{See http://www.mpe.mpg.de/xray/wave/rosat/rra/.} (ROSPSPCCAT/2RXP) using the same observation of \citet{hailey95} provides more reliable coordinates: $\rm{R.A.} = 00^{\rm{h}}02^{\rm{m}}55\fs4$, $\rm{Decl.}=62^{\circ}46'21\farcs0$ (epoch J2000). Adopting this position, the offset between the \xmm\ and \rst\ positions decreases to $4\farcs0$, and that from \tmsrc\ to $4\farcs0$.\\
\indent With the J, H, and K$_{\rm{s}}$ magnitudes of \src/ \tmsrc\ at hand (see \mbox{Section~\ref{ircounterpart}}), we used the relation between the \nh\ of the X-ray best-fits (Table~\ref{fits}) and the interstellar extinction $A_{\rm{V}}$ of \citet[][]{predehl95}, as well as the relations between the extinctions at different wavelengths of \citet{cardelli89} to derive the intrinsic colours of the source \mbox{(J--H)$_{\circ}\simeq0.30$} and (H--K)$_{\circ}\simeq0.06$. In a similar way, taking optical photometric data from the Tycho-2 Catalogue\footnote{See http://www.astro.ku.dk/$\sim$erik/Tycho-2/.} \citep{hog00}, we derived also the colour \mbox{(B--V)$_{\circ}\simeq0.50$}. These values are consistent with a F or G type star in the case of a main-sequence star, or with a G type in the case of a supergiant \citep[e.g.,][]{cox00}. In particular, the intrinsic colours point to a F7-type main-sequence star.\\
\indent In the reasonable frame of a F7-type main-sequence star, the expected absolute optical magnitudes is $M_{\rm{V}}\simeq3.4$ \citep[e.g.,][]{cox00,zombeck07}, implying a distance of \mbox{$\sim$230 pc}. Such a relatively small distance is well consistent with the measure of the photoabsorption derived from the best-fitting models of the X-ray spectrum (\mbox{$N_{\rm H} \simeq 1.5 \times 10^{21}$ cm$^{-2}$}, see Table~\ref{fits}). This value is in fact significantly smaller than the measurements of the interstellar hydrogen in this direction by \citet{dickey90} and \citet{kalberla05}, that give \nh\ values of \mbox{$\sim$(6--$7)\times 10^{21}$ cm$^{-2}$}.
The X-ray-to-optical flux ratio is \mbox{$\log(f_{\rm{X}}/f_{\rm{V}})\simeq -3.3$}, in good agreement with the value of \mbox{$\langle\log(f_{\rm{X}}/f_{\rm{V}})\rangle=-3.7\pm0.7$} obtained by \citet{krautter99} averaging \rst/PSPC and optical data on a sample of 53 F-type stars.
This scenario is further confirmed by the X-ray spectrum of \src\ measured by \xmm\ that is well fit by either the \textsc{APEC} or \textsc{MEKAL} codes, with temperatures typical of non-degenerate stellar atmospheres.\\
\indent Based on the accurately identified counterpart (thanks to \xmm\ imaging capabilities) as well as on the spectral properties of \src, together with the lack of an associated SNR and the absence of X-ray pulsations, we conclude that the source is not a neutron star (in any of its manifestations, including a CCO) and its properties are clearly consistent with a non-degenerate star.

\section*{acknowledgements}
This publication has made use of data products from the Two Micron All Sky Survey, which is a joint project of the University of Massachusetts and the Infrared Processing and Analysis Center/California Institute of Technology, funded by the National Aeronautics and Space Administration and the National Science Foundation.
This research has also made use of the Tycho-2 Catalogue of the 2.5 Million Brightest Stars, of the USNOFS Image and Catalogue Archive operated by the United States Naval Observatory (Flagstaff Station), of HEASARC online services supported by NASA/GSFC, and of the SIMBAD database and the VizieR Catalogue Service operated at CDS (Strasbourg, France). 
The authors acknowledge the support of the Italian Space Agency (contracts ASI/INAF I/023/05/0 and I/008/07/0) and the Italian Ministry for University and Research (grant PRIN 2005 02 5417). 
A.D.L. acknowledges an Italian Space Agency fellowship. 

\bibliographystyle{mn2e}
\bibliography{biblio}
\bsp

\label{lastpage}

\end{document}